\documentclass[11pt,a4paper]{article}
\usepackage{amsmath,amstext,amssymb}
\usepackage{epstopdf}

\def\be{\begin{equation}}
\def\ee{\end{equation}}
\def\ba{\begin{eqnarray}}
\def\ea{\end{eqnarray}}
\newcommand{\psidag}{\psi^{\dag}}

\newcommand{\chidag}{\chi^{\dag}}

\newcommand{\vs}[1]{\vspace{#1 mm}}

\usepackage{graphicx}
\usepackage{dcolumn}
\usepackage{bm}
\usepackage{epsfig}

\usepackage[a4paper, total={6in, 7in}]{geometry}

\begin{document}


\vs{10}
\begin{center}

{\Large \bf Pomeron - Odderon interactions in a reggeon field theory}
\vs{15}

{\large
Jochen Bartels,~\footnote{e-mail address: jochen.bartels@desy.de}$^{a}$
Carlos Contreras~\footnote{e-mail address: carlos.contreras@usm.cl}$^{b}$
and Gian Paolo Vacca~\footnote{e-mail address: vacca@bo.infn.it}$^{c}$
} \\
\vs{10}

$^a${\em II. Institut f\"{u}r Theoretische Physik, Universit\"{a}t Hamburg, Luruper Chaussee 149,
D-22761 Hamburg, Germany}

$^b${\em Departamento de Fisica, Universidad Tecnica Federico Santa Maria, Avda.España 1680, Casilla 110-V, Valparaiso, Chile}

$^c${\em INFN, Sezione di Bologna, DIFA, via Irnerio 46, I-40126 Bologna, Italy}

\vs{15}
{\bf Abstract}
\end{center}
In this paper we extend our recent non perturbative functional renormalization group analysis
of Reggeon Field Theory  to the interactions of Pomeron and Odderon fields. 
We establish the existence of a fixed point and its universal properties,
which exhibits a novel symmetry structure in the space of Odderon-Pomeron interactions.  
As in our previous analysis, this part of our program aims at the investigation of the IR limit of reggeon field theory 
(the limit of high energies and large transverse distances). It should be seen in the broader context 
of trying to connect the nonperturbative infrared region (large transverse distances) with the 
UV region of small transverse distances where the high energy limit of perturbative QCD applies.   
We briefly discuss the implications of our findings for the existence of an Odderon in high energy   
scattering.

\section{Introduction}

In a recent paper \cite{Bartels:2015gou} we have started an analysis of the flow equations
of reggeon field theory (RFT)~\cite{Gribov:1968fg,Gribov:1968uy,AB,Migdal:1973gz,Sugar:1974td,Moshe:1977fe}, 
following the idea that RFT might provide a good description of strong interactions in the Regge limit and infrared region:
rapidity $Y \to \infty$ and transverse distances (impact parameter) $|x_{\perp}| \to \infty$. 
We have used the Wetterich formulation of the functional renormalization group equations~\cite{Wetterich, Morris:1994ie} to study directly the problem in two transverse dimensions.
As our main result we have established the existence of a critical theory (fixed point) with one relevant direction: in the multidimensional space of the parameters of the effective potential,
there exists one direction which is UV attractive (IR repulsive), whereas all other directions are IR attractive. We have verified that the properties of such a fixed point are belonging to the universal behavior which RFT shares with the simplest directed percolation model in statistical physics~\cite{Cardy:1980rr},
and we have found a good agreement with some related numerical MonteCarlo analysis.

This investigation of reggeon field theory should be seen as a first step in searching for an effective theory which describes the high energy Regge limit of QCD. 
Regge theory is being used for analysing the nonperturbative $p\bar{p}$ scattering at the Tevatron (FermiLab), $pp$ scattering at the ISR (CERN), RHIC (BNL) and at the LHC (CERN), and $\gamma p$-scattering at HERA (DESY). On the perturbative side, high energy QCD has been analyzed using 
Regge theory (in particular, the BFKL Pomeron with various applications in $e^+e^-$ scattering, forward jets in $\gamma^*p$ scattering, 
and Mueller-Navelet jets in $pp$ or $p\bar{p}$).
Whereas the first group of high energy scattering processes is characterized by transverse distances 
of hadronic sizes, the second one addresses scattering processes of small transverse diameters.
This suggests to search, in the space of $2+1$ - dimensional reggeon field theories, for an interpolation between the two domains: for long transverse 
distances the Pomeron field has intercept very close to unity and a nonzero $t$-slope,
for short transverse distances the BFKL intercept is significantly above one, and the slope is
very small. 

Within such a program in \cite{Bartels:2015gou} we have restricted our analysis to one reggeon field, the Pomeron field.
Whereas other secondary reggeons (e.g. $\rho$, $\omega$, or $\varphi$) have intercepts well separated from the Pomeron and,
in a first approximation, can therefore safely be neglected, there exists one other Regge singularity for which this is not the case,
the Odderon with intercept at or near one.
In the nonperturbative region, the search for the Odderon has stimulated a long-lasting debate:
the strongest evidence for its existence comes from the observed difference in the dip structure in the $t$-dependence
of elastic cross section of $pp$ or $p\bar{p}$ scattering.  In contrast, in the perturbative region the existence of the Odderon is well-established \cite{Bartels:1999yt}:
in nonabelian  $SU(3)$ gauge theory bound states of reggeized gluons exist for the two Casimir operators, the BFKL Pomeron~\cite{BFKL} and the Odderon. These two states represent the two 
equally important fundamental bound states of the $SU(3)$ gauge theory.    
Whereas the BFKL intercept is well above one, the odderon intercept has been found to be exactly at one. In $N=4$ SUSY theory or in planar QCD, also higher order corrections do not alter this striking feature~\cite{Bartels:2013yga} of the spectrum of the so called BKP equations~\cite{Bartels:1980pe,Kwiecinski:1980wb, Bartels:2012sw}.
Self interactions of the  Pomeron~\cite{Bartels:1994jj, Braun:1997nu, Bartels:2002au,Bartels:2004ef} as well as interactions between Pomeron and Odderon naturally appear in perturbative QCD analysis~\cite{Bartels:1999aw, Bartels:2004hb}.
Analogous results are obtained also in the Color Glass Condensate, dipole and Wilson line 
approach~\cite{Kovchegov:2003dm,Hatta:2005as,Kovner:2005qj}. In summary, the existence of this perturbative Odderon is a manifestation of the $SU(3)$ gauge group of strong interactions: in a $SU(2)$ gauge theory there exist no second Casimir operator, i.e the BFKL Pomeron is the only fundamental bound state of the reggeized gluons. On the other hand, in a $SU(N)$ gauge theory with $N>3$ there are more than two Casimir operators, and one expects more fundamental gluonic bound states.       

The existence of the pertubative region motivates interest in the question whether the Odderon exists also in the nonperturbative region. In analysis which is set up to explore the 
connection between the UV region and the nonpertubative IR region, the Odderon has to be included: the IR fixed point structure should confirm whether the Odderon survives the flow from UV to IR. Also, such an analysis should provide some information on the interaction between Odderon and Pomeron, e.g. on the absorption of the nonperturbative Odderon.           

In this paper we therefore extend our previous analysis to interactions of two fields, Pomeron and Odderon. As we will discuss in more detail below, the fact that the Odderon has odd signature leads to a very pronounced feature of the effective potential in the space of Pomeron-Odderon  field theories. 
As a  first result of our investigations, we establish the existence of a new critical theory (fixed point) which includes both Pomeron and Odderon fields. This fixed point  
now has two relevant directions (plus a possible third relevant direction, for which we need more accurate analysis). The fixed point allows, among others, for a solution where both the Pomeron and the Odderon intercepts at infinite energies approach unity, i.e. in $pp$ scattering an Odderon should exist at high energies.         
Although we still refrain from a quantitative phenomenological analysis of this fixed point,
we nevertheless can deduce a few predictions for the Pomeron-Odderon sector at high 
energies. A more complete analysis of the phase diagram, in particular the search for
the possibility of other fixed points, requires further investigations and will be left for a future publication.  

Again this investigation can have implications in the statistical physics of generalized multicomponent directed percolation models. In absence of specific symmetries these models are usually considered to belong to a single universality class. We find first evidence that the RFT, which should be related to the infrared high energy limit of QCD, 
is instead characterized by symmetry properties which map it on a novel percolating system.

This paper is organized as follows. In section 2 we describe our setup. The calculations
of the fixed point conditions ($\beta$-functions of the set of the parameters of the effective 
potential) represent a rather nontrivial extension of the pure Pomeron case. 
In the following section we present and discuss numerical results. In a concluding section we discuss first implication for real physics.
In an appendix we briefly discuss, for future purposes, stationary points of the combined Pomeron-Odderon effective potential.        

\section{The setup}

In the following we consider interactions between Pomeron and Odderon fields.
As before, $\psi,\psidag$ denote the Pomeron field, and for the Odderon we introduce
the field $\chi,\chidag$. The effective action has the form: 
\ba
\Gamma[\psidag,\psi,\chidag,\chi]&=&\int \!  \, \mathrm{d}^D x \,  \mathrm{d} \tau
 \left( Z_{P }(\frac{1}{2} \psi^{\dag} \overset\leftrightarrow{\partial}_{\! \tau} \psi -\alpha'_{P } \psidag\nabla^{2}\psi)
+ Z_{O }(\frac{1}{2}\chidag\overset\leftrightarrow{\partial}_{\! \tau}\chi -\alpha'_{O } \chidag\nabla^{2}\chi)\right.\nonumber\\ 
&& \left.+ V_k[\psi,\psidag,\chi,\chidag] \right).
\ea
Here $D$ denotes the number of spatial dimension. $D=2$ is the physical case of our interest, but in our analytic formulae (Section $2$ and Sections $3.1$-$3$) we find it convenient to keep $D$ as a continuous parameter. 
The particular case $D=4$ is the scaling dimension (critical dimension) of reggeon field theory, and later on (Section $3.4$)  it will be useful to refer to results obtained from the $\epsilon$ expansion in $D=4-\epsilon$ dimensions. The numerical results of our fixed point analysis (section 4.1) are obtained for $D=2$. To illustrate the quality of our approximation we find it instructive (section 4.2) to compute anomalous dimensions in the whole interval $0<D<4$. 

For the lowest truncation the effective action takes  the form:
\ba
V_3&=&-\mu_P \psidag \psi +i\lambda \psidag (\psi +\psidag) \psi -\nonumber\\
&&  -\mu_O \chidag \chi +i\lambda_2 \chidag (\psi +\psidag) \chi + \lambda_3 
(\psidag \chi^2 + {\chidag}^2 \psi).
\ea
For the quartic truncation we add the following terms:
\ba
V_4&=&\lambda_{41}(\psi \psidag)^2+
\lambda_{42} \psi \psidag (\psi^2+{\psidag}^2) +\lambda_{43}(\chi\chidag)^2 
+i \lambda_{44} \chi\chidag (\chi^2+{\chidag}^2)\nonumber\\
&&+i \lambda_{45}\psi \psidag(\chi^2 
+{\chidag}^2) +\lambda_{46}\psi\psidag \chi \chidag 
+\lambda_{47}\chi \chidag(\psi^2+{\psidag}^2).
\ea 
Similarly, the quintic truncation has the following eleven terms:
\ba
V_5&=& i\left(\lambda_{51} (\psi \psidag)^2 (\psi+\psidag)+\lambda_{52} \, \psi \psidag (\psi^3+{\psidag}^3)
+\lambda_{53} \, \chi \chidag (\psi^3+{\psidag}^3)
+\lambda_{54} \,\psi \psidag \chi \chidag(\psi+\psidag) \right)\nonumber\\
&&+ \lambda_{55} (\chi^2 {\psidag}^3 + {\chidag}^2 \psi^3)
+\lambda_{56}\, (\chi^2 {\psidag}^2 \psi + {\chidag}^2 \psidag \psi^2)
+ \lambda_{57}\,(\chi^2 \psidag \psi^2 + {\chidag}^2 {\psidag}^2 \psi)\nonumber\\
&&+i\left(\lambda_{58}\,( \chi^4 \psidag + {\chidag}^4 \psi)
+\lambda_{59} \, (\chi \chidag)^2 (\psi+\psidag)\right) \nonumber\\
&&+\lambda_{510} \,\chi \chidag (\chi^2 \psi + {\chidag}^2 \psidag)
+\lambda_{511} \, \chi \chidag (\chi^2 \psidag + {\chidag}^2 \psi).
\ea
It is important to note the differences in the structure of the effective potential 
compared to the pure Pomeron case.
As described in \cite{Bartels:2015gou}, for the pure Pomeron case the couplings are real-valued for even powers of the Pomeron fields, whereas odd powers require imaginary couplings. This is a consequence of the even-signature of the Pomeron exchange which leads to special trigonometric factors in front of multi-pomeron cut contributions  in the $t$-channel unitarity equations: the $n$-Pomeron contribution comes with a factor $(-1)^{n-1}$. 
This means, in particular, that the two Pomeron cut contribution to the Pomeron self-energy has a minus sign which is obtained by requiring the triple Pomeron 
coupling to be purely imaginary. 

For the Odderon the situation is different: the Odderon has negative signature. This has several consequences. First, because of signature conservation, $t$-channel states with an odd number of Odderons never mix with pure Pomeron channels. Second, the transition 
$P \to OO$ is real valued: the two-Odderon cut is positive (in contrast to the two Pomeron cut), and there is no need for an imaginary coupling. On the other hand, the transition $O \to OP$ has to be imaginary, since the Odderon-Pomeron cut carries a minus sign. As a result, all triple couplings are imaginary, except for the real-valued transition $ P\to OO$.

In the sector of quartic couplings, all couplings involving Pomerons only are real-valued.
Once the Odderon is included, again most quartic couplings remain real, but there are two 
exceptions: the transitions $O \to OOO$ and $P \to P+OO$  are imaginary.
This can be easily understood considering a contribution to such quartic vertices coming by the composition of two triple ones.
For the quintic part the 'exceptional' terms are in the second and fourth lines: in all
these terms we either create or annihilate a pair of Odderons.  

The signature-conservation rule, together with the appearance of these 'exceptional' cases suggests the following transition rules:\\
(i) states with even and odd numbers of Odderon never mix.\\ 
(ii) states will be labelled by the number of Odderon pairs, $n$. We assign a quantum number $O_n$.
Signature rules imply that transitions changing $n$ by odd numbers come with 'exceptional' couplings (e.g the transitions $P \to OO$,
$O \to OOO$, or $P \to P+OO$), whereas transitions changing $n$ by even numbers are 'normal' and have the same structure as pure Pomeron couplings (e.g., the transition: Pomeron $to$ four Odderons is imaginary).\\
This suggests to decompose the effective potential into a sum 
terms $V^{(n)}$:
\be
\label{pot-symmetry}
V=V^{\Delta n=0} +   \Delta V^{|\Delta n|=1} + \Delta V^{|\Delta n|=2}+...
\ee 
where the first term conserves $n$, the number of odderon pairs, the second one 
changes $n$ by one etc. 

In the perturbative region, the transition $P \to OO$ has been computed \cite{Bartels:1999aw,Bartels:2004hb} and found to be nonzero.
As one of our results we shall see that the dynamics allows for a critical theory
(as a fixed point of the flow in the local potential approximation (LPA), eventually including anomalous dimensions (LPA') ) at which $n$ is conserved,
i.e. all couplings which change $O_n$ go to zero:
\be
 \Delta V^{|\Delta n|=1} \to 0, \, \, \, \, \Delta V^{|\Delta n|=2} \to 0,....  
\ee   
This applies, in particular, the coupling of the $P \to OO$ transition.

Next we introduce dimensionless variables. The field variables are rescaled as follows:
\be
\tilde{\psi} = Z_P^{1/2} k^{-D/2} \psi, \,\, \,\, \tilde{\chi}= Z_O^{1/2} k^{-D/2} \chi.
\label{dimlessf}
\ee
For the potential we introduce
\be
\tilde{V}=\frac{V}{\alpha'_P k^{D+2}}.
\label{dimlessv}
\ee
This choice implies that we introduce the dimensionless ratio 
\be
r=\frac{\alpha'_O}{\alpha'_P},
\ee
and the Odderon slope $\alpha'_O$ will be written as 
\be
\label{ratio-r}
\alpha'_O= r \,\alpha'_P.
\ee  
Finally, using Eq.~\eqref{dimlessf} and~\eqref{dimlessv}, the couplings are rescaled in the following way:
\ba
\label{dimensional}
&&\tilde{\mu}_P=\frac{\mu_P}{Z_P \alpha'_P k^2}, \,\,\,\,\tilde{\mu}_O=\frac{\mu_O}{Z_O \alpha'_P k^2}, \nonumber\\ 
&&\tilde{\lambda}=\frac{\lambda}{Z_P^{3/2} \alpha'_P k^2} k^{D/2},\,\, \,\,
\tilde{\lambda}_{2,3}= \frac{\lambda_{2,3}}{Z_O Z_P^{1/2} \alpha'_Pk^2} k^{D/2}.
\ea
With these definitions the classical scaling (canonical) of the potential which would result by neglecting the quantum fluctuations in the flow equation becomes:
\be
\left(-(D+2) + \zeta_P\right) \tilde{V} + \left(\frac{D}{2} + \frac{\eta_P}{2} \right)
\left( \tilde{\psi} \frac{\partial \tilde{V}}{\partial \tilde{\psi}} + \tilde{\psidag} \frac{\partial \tilde{V}}{\partial \tilde{\psidag}} \right)
+ \left( \frac{D}{2} + \frac{\eta_O}{2} \right) \left(\tilde{\chi} \frac{\partial \tilde{V}}{\partial \tilde{\chi}} + \tilde{\chidag} \frac{\partial \tilde{V}}{\partial \tilde{\chidag}} \right).
\ee
The scale $k$ dependent regulator functions are chosen as follows:
\ba
&&R_P(q^2)=Z_P \alpha'_P (k^2-q^2) \Theta(k^2-q^2),\nonumber\\
&&R_O(q^2)=Z_O \alpha'_O (k^2-q^2) \Theta(k^2-q^2)=r Z_O \alpha'_P (k^2-q^2) \Theta(k^2-q^2).
\ea
This optimized regulator~\cite{Litim} allows for a simple analytic integration in a closed form.
Moreover we define the anomalous dimensions: 
\be
\label{anomalous-eta}
\eta_P= - \frac{1}{Z_P} \partial_t Z_P,\,\, \,\,   \eta_O= - \frac{1}{Z_O} \partial_t Z_O
\ee
and
\be
\label{anomalous-zeta}
\zeta_P=-\frac{1}{\alpha'_P} \partial_t \alpha'_P,\,\, \,\, \zeta_O=-\frac{1}{\alpha'_O} \partial_t \alpha'_O.
\ee

\section{RG flow}
\subsection{Flow equations}
In order to find the flow equation of the potential (which included Pomeron and Odderon intercepts (masses) as well as all possible interactions)
we need to compute the rhs of the dimensionful flow equations which result from scale $k$ controlled contributions from quantum fluctuations:
\be
\partial_{t}\Gamma=\frac{1}{2}Tr[\Gamma^{(2)}+\mathbb{R}]^{-1}
\partial_{t}\mathbb{R}.
\label{eq:exactflow1}
\ee
The trace on the rhs extends over a $4x4$ matrix. The propagator matrix can be written the following form:
\ba
\Gamma^{(2)} +\mathbb{R} = \left( \begin{array}{cc} \Gamma^{(2)}_P & \Gamma_{PO} \\
\Gamma_{OP} & \Gamma^{(2)}_O\end{array} \right),
\label{propagator}
\ea
where the $2x2$ block matrices are:
\ba
\Gamma^{(2)}_P=  \left( \begin{array}{cc} V_{\psi \psi} &  Z_P( -i\omega  +  \alpha'_P q^2) +R_P + V_{\psi \psidag} \\  Z_P( i \omega +  \alpha'_P q^2) +R_P + V_{\psidag \psi} & V_{\psidag \psidag}
\end{array} \right) ,
\ea
\ba
\Gamma^{(2)}_O=  \left( \begin{array}{cc} V_{\chi \chi} &  Z_O( -i\omega  +  \alpha'_O q^2) +R_O + V_{\chi \chidag} \\  Z_O( i \omega +  \alpha'_O q^2) +R_O + V_{\chidag \chi} & V_{\chidag \chidag}
\end{array} \right) ,
\ea
\ba
\Gamma_{PO} = \left( \begin{array}{cc} V_{\psi\chi}& V_{\psi \chidag}\\
 V_{\psidag\chi}& V_{\psidag \chidag} \end{array}\right),
\ea
and  
\ba
\Gamma_{OP} = \left( \begin{array}{cc} V_{\chi\psi}& V_{\chi \psidag}\\
 V_{\chidag\psi}& V_{\psi \psidag} \end{array}\right).
\ea  

The momentum integral contained in the trace can be done in the same way as described in 
\cite{Bartels:2015gou}. The energy integral will be performed by complex integration.
Unfortunately the analytic expression for the full flow of the potential is quite involved and difficult to handle.
Since we are interested in an analysis based on polynomial expansions of the potential in terms of the Pomeron and Odderon fields,
we find it more convenient to derive directly the flow equations for the polynomial coefficients (couplings).

In this work we shall limit ourself in analyzing the flow of the potential expanded around the origin (zero fields), i.e. we shall employ a weak field expansion.
We shall consider more refined analysis in a future investigation.
Therefore, for the  derivation of the beta-functions of the couplings we find it convenient to expand the inverse of   
(\ref{propagator}) in the following way:
\ba
[\Gamma^{(2)} +\mathbb{R}]^{-1}&=&[\Gamma^{(2)}_{free} -V_{int}]^{-1}\nonumber\\
& =&G(\omega,q)+G(\omega,q) V_{int} G(\omega,q)+G(\omega,q) V_{int} G(\omega,q)V_{int} G(\omega,q)+ ...
\ea
Here we absorb the masses (intercepts minus one) into the free propagators:
\ba
\label{free-prop}
G(\omega,q) = \left( \begin{array}{cc} G_P(\omega,q) & 0\\ 0& G_O(\omega,q)
\end{array} \right),
\ea
where 
\ba
G_P(\omega,q)=\left( \begin{array}{cc} 0& (Z_P( -i\omega  +  \alpha'_P q^2) +R_P-\mu_P)^{-1} \\(Z_P( i\omega  +  \alpha'_P q^2) +R_P-\mu_P)^{-1}& 0 \end{array}
\right)
\ea
and
\ba
G_O(\omega,q)=\left( \begin{array}{cc} 0& (Z_O( -i\omega  +  \alpha'_O q^2) +R_O-\mu_O)^{-1} \\(Z_O( i\omega  +  \alpha'_O q^2) +R_O-\mu_O)^{-1}& 0 \end{array}
\right).
\ea
The interaction matrix $V_{int}$ is derived  from the effective potential, after removal of the reggeon masses:
\ba
V_{int}=- \left( \begin{array}{cccc}V^{r}_{\psi\psi}&V^{r}_{\psi\psidag}&V^{r}_{\psi\chi}&V^{r}_{\psi\chidag}\\
V^{r}_{\psidag\psi}&V^{r}_{\psidag\psidag}&V^{r}_{\psidag\chi}&V^{r}_{\psidag\chidag}\\
V^{r}_{\chi\psi}&V^{r}_{\chi\psidag}&V^{r}_{\chi\chi}&V^{r}_{\chi\chidag}\\
V^{r}_{\chidag\psi}&V^{r}_{\chidag\psidag}&V^{r}_{\chidag\chi}&V^{r}_{\chidag\chidag} \end{array} \right).
\ea
Here the upper script 'r' reminds that the reggeon masses have been removed. 

Finally we define the regulator matrix consisting of two block matrices.
First we define
\ba
O_{\pm}= \left( \begin{array}{cc} 0 & 1 \\  \pm 1& 0
\end{array} \right).
\ea
With this we find
\ba
\dot{R}=\left( \begin{array}{cc} \dot{R}_{P}&0\\0&\dot{R}_{O} \end{array} \right),
\ea
where 
\ba
\dot{R}_P= \partial_t R_P(q^2)O_{+}
\ea
and 
\ba
\dot{R}_O= \partial_t R_O(q^2) O_{+}.
\ea
After the momentum integrals and after the use of dimensionless variables, the factors of the block matrices can be replaced by   
\ba
\dot{R}_P \to  N_D A_D(\eta_P,\zeta_P) O_{+}
\ea
and
\ba
\dot{R}_O \to r N_D A_D(\eta_O,\zeta_O) O_{+},
\ea
where the factors $N_D$ and $A_D$ are defined in \cite{Bartels:2015gou}, and 
$\eta_P,\zeta_P$ and $\eta_O,\zeta_O$ are the anomalous dimensions of the Pomeron and 
Odderon fields, respectively. 

We are this left with the energy integrals in the expansion:
\be
\label{trace-expansion} 
\int \frac{dz'}{2\pi} Tr \Big[ R \,\, G(z')\left(1 +   V_{int} G(z') + 
  V_{int} G(z') V_{int} G(z')+  V_{int} G(z') V_{int} G(z') V_{int} G(z')
+
...\right)\Big] 
\ee
where $z'=i\omega'/(\alpha'_P k^2)$, and the free propagators in (\ref{free-prop}), as a result of the momentum integration and the use of dimensionless variables, have become:
\ba
G_P(\omega,q) \to G_P(z) =\left( \begin{array}{cc} 0& (-z  + 1-\tilde{\mu}_P)^{-1} \\(z +1
 -\tilde{\mu}_P)^{-1}& 0 \end{array}
\right),
\ea
\ba
G_O(\omega,q) \to G_O(z) =\left( \begin{array}{cc} 0& (-z  + r-\tilde{\mu}_O)^{-1} \\(z +r
 -\tilde{\mu}_O)^{-1}& 0 \end{array}
\right).
\ea
For the derivation of the beta functions, we take derivatives of (\ref {trace-expansion}) with respect to the field variables and subsequently set the fields equal to zero.
The first two terms on the rhs does not contribute, and depending upon the truncation of the potential, 
we only need a finite number of terms. 

\subsection{$\beta$ functions}

We begin with the lowest (cubic) truncation. For this approximation of the effective potential, we keep on the rhs of (\ref {trace-expansion}) the terms with two and three V's.
The $z$-integral is done by complex integration. We report here the result for the region $r-\mu_O >0$ which can be verified "a posteriori" to be the physical relevant region.
The beta functions in the complementary region $r-\mu_O <0$ can be computed in a similar way but we shall not discuss them further.
Including also the canonical part on the rhs of the flow equations we find: 
\ba
\dot{\mu}_P &=&   (-2 +\eta_P+\zeta_P) \mu_P
+ 2 A_P \frac{\lambda^2}{(1-\mu_P)^2} -  2 A_O r \frac{\lambda_3^2}{(r-\mu_O)^2} \nonumber\\
\dot{\mu}_O &=&   (-2 +\eta_O+\zeta_P) \mu_O +2 (A_P + A_O r) \frac{\lambda_2^2}{(1+r-\mu_P-\mu_O)^2} 
\nonumber\\
\dot{\lambda}&=&(-2+D/2+\zeta_P +\frac{3}{2} \eta_P)\lambda
+8A_P  \frac{\lambda^3}{(1-\mu_P)^3 }- 
4A_O r  \frac{\lambda_2\lambda_3^2}{(r-\mu_O)^3 }
\nonumber\\
\dot{\lambda}_2&=&(-2+D/2+\zeta_P +\frac{1}{2}\eta_P +\eta_O)\lambda_2\nonumber\\
&& + \frac{2\lambda \lambda_2^2(6A_P +5A_O r)+4\lambda_2^3 (A_P+A_O r)-4\lambda_2\lambda_3^2 (A_P+2A_Or)}{(1+r-\mu_P-\mu_O)^3}
  \nonumber\\
&&+\frac{2A_P\lambda\lambda_2^2(r-\mu_O)^2}{(1-\mu_P)^2 (1+r-\mu_P-\mu_O)^3}
-\frac{4A_Or\lambda_2\lambda_3^2(1-\mu_P)^2}{(1-\mu_O)^2 (1+r-\mu_P-\mu_O)^3}
\nonumber\\
&&+\frac{2\lambda\lambda_2^2(3A_P+A_Or)(r-\mu_O) }{(1-\mu_P) (1+r-\mu_P-\mu_O)^3}
-\frac{4\lambda_2\lambda_3^2(A_P+3A_Or)(1-\mu_P) }{(r-\mu_O) (1+r-\mu_P-\mu_O)^3}
\nonumber\\
\dot{\lambda}_3&=&(-2+D/2+\zeta_P +\frac{1}{2}\eta_P +\eta_O)\lambda_3 \nonumber\\
&&+\frac{2\lambda_2^2\lambda_3(A_P+2A_Or)}{(r-\mu_O)(1+r-\mu_P-\mu_O)^2}
+\frac{4\lambda\lambda_2\lambda_3(2A_P+A_Or)}{(1-\mu_P)(1+r-\mu_P-\mu_O)^2}
\nonumber\\
&&+\frac{2\lambda_2^2\lambda_3 A_Or(1-\mu_P)}{(r-\mu_O)^2(1+r-\mu_P-\mu_O)^2}
+\frac{4\lambda\lambda_2\lambda_3 A_P(r-\mu_O)}{(1-\mu_P)^2(1+r-\mu_P-\mu_O)^2}.
\label{beta3}
 \ea
Here we have defined 
\be
A_P=N_D A_D(\eta_P,\zeta_P),  \,\,\,A_O=N_D A_D(\eta_O,\zeta_O).
\label{A_PO}
\ee
For the next truncation, the quartic approximation, we have to retain also the next term on the rhs of (\ref {trace-expansion}) (containing four factors of $V_{int}$).
The results for the beta functions are already lengthy and will not be listed here. For the truncations of fourth order and beyond we have used symbolic computational tools (Mathematica).
\subsection{Anomalous dimensions}

Having derived the beta function we need to mention a novel feature which was not present for the pure Pomeron case: all beta functions will depend upon the parameter $r$ defined in (\ref{ratio-r}), the ratio of the Odderon and Pomeron slopes. This dimensionless quantity by itself depends upon the cutoff parameter $k$ and therefore has its own beta function.
The critical theory satisfies the fixed point condition $\dot{r}=0$.
We therefore need not only the beta functions for the parameters of the effective potential (coupling constants) but also the anomalous dimensions.
With the anomalous dimensions defined in (\ref{anomalous-eta}), the evolution equation for $r$ then becomes:
\be
\label{rdot}
\dot{r} = r \left(- \zeta_O+ \zeta_P \right),
\ee 
which tells that at criticality the Pomeron and Odderon transverse space scaling laws do coincide.

In order to obtain the anomalous dimensions we first define the two-point vertex functions:
\be
\Gamma^{(1,1)}_P (\omega,q) = \frac{\delta^2 \Gamma}{\delta \psi(\omega,q) \delta \psidag(\omega,q)}|_{\psi=\psidag=\chi=\chidag=0}
\ee
and
\be
\Gamma^{(1,1)}_O (\omega,q)= \frac{\delta^2 \Gamma}{\delta \chi(\omega,q) \delta \chidag(\omega,q)}|_{\psi=\psidag=\chi=\chidag=0}.
\ee
From the flow equations we obtain:
\ba
\dot{\Gamma}^{(1,1)}_P(\omega,q) &=&\alpha'_P \int \frac{d z'd^Dq'}{(2\pi)^{D+1}} Tr\left [ \dot{R} \, G(z',q') \frac{\delta V_{int}}{\delta \psidag} G(z+z',q+q') \frac{\delta V_{int}}{\delta \psi}
G(z',q') \right]|_{O}+...\nonumber \\ \\
\dot{\Gamma}^{(1,1)}_O(\omega,q)&=&\alpha'_P \int \frac{d z'd^Dq'}{(2\pi)^{D+1}} Tr\left[ \dot{R} \, G(z',q') \frac{\delta V_{int}}{\delta \chidag} G(z+z',q+q') \frac{\delta V_{int}}{\delta \chi}
G(z',q') \right]|_{O} +...\nonumber\\  
\ea
where the subscript 'O' indicates that, after differentiation, we have set all field variables
inside the trace equal to zero:  $\psi=\psidag=\chi=\chidag=0$, and the dots indicate that there are more 
terms containing second derivatives of $V_{int}$ with respect to the field variables
which will not contribute when taking derivatives in $z'$ or ${q'}^2$. We have already 
taken into account that, from the derivatives with respect to $\psi,\psidag$ (or $\chi,
\chidag$)  we have two identical contributions which compensate the overall factor $\frac{1}{2}$.    
The anomalous dimensions are obtained by taking derivatives with respect to energy and momentum:
\ba
Z_P&=& \lim_{\omega\to 0,q\to 0} \frac{\partial}{\partial (i\omega)}  \Gamma^{(1,1)}_P (\omega,q)\\
Z_O&=& \lim_{\omega\to 0,q\to 0} \frac{\partial}{\partial (i\omega)}  \Gamma^{(1,1)}_O (\omega,q)
\ea
and
\ba
Z_P\alpha'_P &=& \lim_{\omega\to 0,q\to 0} \frac{\partial}{\partial q^2}  \Gamma^{(1,1)}_P (\omega,q)\\
Z_O \alpha'_O &=& \lim_{\omega\to 0,q\to 0} \frac{\partial}{\partial q^2}  \Gamma^{(1,1)}_O (\omega,q).
\ea
We introduce 
\ba
\dot{\Gamma}^{(1,1)}_P &=& I^{(1,1)}_P(\omega,q)\\
\dot{\Gamma}^{(1,1)}_O &=& I^{(1,1)}_O(\omega,q).
\ea
The anomalous dimensions are then given by:
\ba
-\eta_P &=& \frac{1}{Z_P} \lim_{\omega\to 0, q\to 0}  \frac{\partial}{\partial (i\omega)} I^{(1,1)}_P(\omega,q)\\
-\eta_O &=& \frac{1}{Z_O} \lim_{\omega\to 0, q\to 0}  \frac{\partial}{\partial (i\omega)} I^{(1,1)}_O(\omega,q)
\ea
and
\ba
-\eta_P - \zeta_P &=& \frac{1}{Z_P \alpha'_P} \lim_{\omega\to 0, q\to 0} \frac{\partial}{\partial q^2} I^{(1,1)}_P(\omega,q)\\
-\eta_O - \zeta_O &=& \frac{1}{Z_O \alpha'_O} \lim_{\omega\to 0, q\to 0} \frac{\partial}{\partial q^2} I^{(1,1)}_O(\omega,q).
\ea

The calculation of the derivatives with respect to $z$ and $q^2$ has been described in  
\cite{Bartels:2015gou}. For the $z$-derivative we obtain after the momentum integral:
\ba
&&\frac{1}{Z_P\alpha'_P} \frac{d I^{(11)}_P}{dz} = 2 N_D \int \frac{ dz'}{2\pi i} \cdot\\&&\cdot Tr\left[ \left( \begin{array}{cc} A_D(\eta_P,\zeta_P) O_{+} & 0 \\ 0 & r A_D(\eta_O,\zeta_O) O_{+}\end{array} \right)
G(z')\frac {\delta V_{int}}{\delta \psidag}G(z') \left( \begin{array}{cc}   O_{-} & 0\\0& O_{-}\end{array} \right)G(z')\frac{ \delta V_{int}}{\delta \psi} G(z')
 \right]\nonumber
\ea  
\ba
&&\frac{1}{Z_O\alpha'_P}\frac{d I^{(11)}_O}{dz} = 2 N_D \int \frac{ dz'}{2\pi i} \cdot\\&&\cdot Tr\left[ \left( \begin{array}{cc} A_D(\eta_P,\zeta_P) O_{+} & 0 \\ 0 & r A_D(\eta_O,\zeta_O) O_{+}\end{array} \right)
G(z')\frac {\delta V_{int}}{\delta \chidag}G(z') \left( \begin{array}{cc}   O_{-} & 0\\0& O_{-}\end{array} \right)G(z')\frac{ \delta V_{int}}{\delta \chi} G(z')
 \right].\nonumber
\ea  
Similarly, for the $q^2$ derivative we find:
\ba
\frac{1}{Z_P\alpha'_P} \frac{d I^{(11)}_P}{dq^2} = \frac{N_D}{D}  \int \frac{ dz'}{2\pi i}  Tr\left[ \left( \begin{array}{cc}  O_{+} & 0 \\ 0 & r O_{+}\end{array} \right)
G(z')\frac {\delta V_{int}}{\delta \psidag}G(z') \left( \begin{array}{cc}   O_{+} & 0\\0& r O_{+}\end{array} \right)G(z')\frac{ \delta V_{int}}{\delta \psi} G(z')
 \right]\nonumber\\
\ea  
\ba
\frac{1}{Z_O\alpha'_O}\frac{d I^{(11)}_O}{dq^2} =  \frac{N_D}{ rD} \int \frac{ dz'}{2\pi i}  Tr\left[ \left( \begin{array}{cc} O_{+} & 0 \\ 0 & r O_{+}\end{array} \right)
G(z')\frac {\delta V_{int}}{\delta \chidag}G(z') \left( \begin{array}{cc}   O_{+} & 0\\0& r O_{+}\end{array} \right)G(z')\frac{ \delta V_{int}}{\delta \chi} G(z')
 \right].\nonumber\\
\ea  

We quote the results for the expansion around zero fields (in this case, the results do not depend upon the truncation since only cubic couplings are involved):
\be
\eta_P=-\frac{2 A_P \lambda^2}{(1-\mu_P)^3} + \frac{2 A_O r \lambda_3^2}{(r-\mu_O)^3}
\label{anom1}
\ee
\be
\eta_O=- \frac{4(A_P + A_O\, r)\lambda_2^2}{(1+r-\mu_P-\mu_O)^3}
\ee
 and
\be
\eta_P+\zeta_P=-\frac{N_D \lambda^2}{D(1-\mu_P)^3} + \frac{N_D r^2 \lambda_3^2}{D (r-\mu_O)^3}
\ee
\be
\eta_O+\zeta_O=- \frac{4N_D \lambda_2^2}{D(1+r-\mu_P-\mu_O)^3}.
\label{anom4}
\ee  

\subsection{Analysis near $D=4$: $\epsilon$-expansion}

From a quick look at the beta functions given in Eq.~\eqref{beta3} of the couplings $\lambda$, $\lambda_2$  and $\lambda_3$ of the cubic truncation one sees that they do not scale when $D\to 4$, which is indeed the scaling (critical) dimension of the system.
In this section we show the results of an analysis of the theory close to the critical dimension ($D=4-\epsilon$) at one loop, restricted to the cubic truncation only.
Such an analysis can help to identify a possible critical behavior of the system which may survive, at a qualitative level, down to $D=2$.
In the next Section, after a numerical analysis with higher truncations in $D=2$, we shall also investigate numerically the fixed points of the cubic truncation for continuos dimensions ($0<D<4$).

Evaluating the Eq.~\eqref{beta3},~\eqref{A_PO} and~\eqref{anom1}-\eqref{anom4} for $D=4-\epsilon$ one searches for solutions such that $\lambda^2, \lambda_2^2,  \lambda_3^2, \mu_P, \mu_O=O(\epsilon)$. We find that, apart from the pure Pomeron scaling solution, in the presence of the Odderon field only other fixed point is allowed:
\ba
\label{specialFPepsilon}
&{}&\!\!\!\!\mu_P=\frac{\epsilon}{12},\quad \lambda^2=\frac{8\pi^2}{3} \epsilon, \quad \eta_P=-\frac{\epsilon}{6}, \quad \zeta_P=\zeta_O=\frac{\epsilon}{12}, \\
&{}&\!\!\!\!\mu_O=\frac{95\!+\!17\sqrt{33}}{2304}\epsilon, \quad \lambda_2^2=\frac{23\sqrt{6}\!+\!11\sqrt{22}}{48}\epsilon, \quad \lambda_3=0,
 \quad \eta_O=-\frac{7\!+\!\sqrt{33}}{72}\epsilon,
\quad r=\frac{3}{16}(\sqrt{33}\!-\!1).\nonumber
\ea

Moreover the spectral analysis of the stability matrix is able to show the other universal quantities of the system, apart from the anomalous dimensions.
In particular we find two negative eigenvalues, associated to two relevant directions, and the corresponding critical exponents:

\ba
\lambda^{(1)}&=&-2+\frac{\epsilon}{4} \,\,\,\rightarrow \nu_P=\frac{1}{2} +\frac{\epsilon}{16}\nonumber \\
\lambda^{(2)}&=&-2+\frac{\epsilon}{12} \rightarrow \nu_O=\frac{1}{2} +\frac{\epsilon}{48}.
\label{nuepsilon}
\ea

We note that the most negative eigenvalue (strongest relevat operator) is associated to the Odderon sector.
We have not found other solutions with all real couplings and $\lambda_3\ne 0$.
We also note that the values of the couplings and the critical exponents and anomalous dimensions in the Pomeron sector are exactly the same as in the pure pomeron case~\cite{Bartels:2015gou}.
This seems to favour, at least in the vicinity of the critical dimension $D=4$, the existence of just two non trivial fixed points, one with the pure pomeron content, and another one with both interacting fields,
where the interaction responsible for creating the odderon fields is turned off. 
That is the scaling solution of Eq.~\eqref{specialFPepsilon} is a theory conserving the Odderon number, and the direction in parameter space which contains the operator breaking such conservation is irrelevant.
\section{Numerical results}
\label{numerics}
\subsection{Search for fixed points}
Let us now focus on the physical case of $D=2$ transverse dimensions.
In a first step of searching fixed point theories (scaling solutions)\footnote{We stress that such solutions for the fixed point of the flow cannot be related to a CFT in the whole $2+1$ dimensional
space because they are characterized by anisotropic scaling between the rapidity direction and the transverse space.} we set the anomalous dimensions equal to zero and search for fixed points of the whole potential, which should be defined on the whole field space.

As already said, in this preliminary investigation we perform a weak field expansion and consider the beta-functions of the corresponding couplings,
complemented by the equation for $\dot{r}$, (\ref{rdot}).
In our analysis we search in the region $r-\mu_O >0$. The computation of the beta functions depends on this condition and we have also computed them in the case of $r-\mu_O\le 0$.
We have performed a fixed point analysis in both regions and found that only in the first one we find physically relevant solutions.

Our analysis is essentially in the LPA approximation with the addition of an extra coupling $r$,
depending on the anomalous dimensions $\zeta_P$ and $\zeta_O$, which we have evaluated at the lowest order.
In all the other beta functions the anomalous dimensions have been completely neglected.
Such a strategy is based on our previous experience with the RFT analysis of the pure Pomeron theory. 
Indeed, the polynomial expansion around the origin without the inclusion of the anomalous dimensions was giving, 
even if the convergence was slow, a very good estimate of the critical exponent $\nu$.
Instead the inclusion of the anomalous dimensions has been shown to be reliable only in the lowest cubic truncation, while it was giving completely incorrect results in higher truncations. 
In order to include the anomalous dimensions we were choosing a different (stationary) point of expansion for the potential. 
We shall not do this here and leave it for future analysis.
In this paper we report on the existence fixed point which looks most promising.  

In the cubic truncation we find a fixed point solution with the following values: 
$(\mu_P= 0.111111, \mu_O=  0.110753, \lambda=1.05034, 
\lambda_2= 1.44665, \lambda_3= 0, r =0.921810)$.
This fixed point has three negative and three positive eigenvalues, i.e. there are three relevant directions. 
Two of the negative eigenvalues $\lambda_O=-1.9398$ and  $\lambda_P=-1.8860$ are associated to the $\nu_P$ and $\nu_O$ critical exponents, respectively.
The third negative eigenvalue $\lambda^{(3)}=-0.0916$ is close to zero and with an eigenvector mainly associated to the $r$ coupling.
Since this is associated to the anomalous dimensions which are evaluated at the lowest order one should take the possibility of having a third relavant direction "cum grano salis"
Clearly further checks with more accurate analysis are needed.

This solution, where the Pomeron sector is the same as in the pure Pomeron case, but the Odderon sector is nontrivial, is also the one found in the $\epsilon$-expansion analysis close to $D-4$.
We observe a  'decoupling' of the two sectors : compared to the pure Pomeron case, the Pomeron is not affected by the presence of the Odderon, whereas the Odderon 'feels' the Pomeron. 
This decoupling is due to the vanishing of the exceptional triple coupling $\lambda_3$, i.e. the vanishing of the first symmetry breaking term, $\Delta V$ on the rhs of  (\ref{pot-symmetry}). We remind that in the UV region where perturbative QCD applies, the coupling
$P \to OO$ is nonzero \cite{Bartels:1999aw}. 
There is also an interesting property of the eigenvectors 
at this fixed point: there is one eigenvector which points in the direction of the 'exceptional' coupling $\lambda_3$ and is orthogonal to all the other eigenvectors (defining both relevant or irrelevant directions).
It has a positive eigenvalue (so that the corresponding interaction is irrelevant) and thus lies inside the critical subspace. 

All these features also appear in the following solution obtained in the quartic truncation:
($\mu_P=0.274381$, $\mu_O=0.26979$, $\lambda=1.34738$, 
$\lambda_2=  1.79401$, $\lambda_3= 0.$, $\lambda_{41}= -2.88712$,
$\lambda_{42}= -1.27076$, $\lambda_{43}= -0.83228$, $\lambda_{44}= 0.$, 
$\lambda_{45}= 0.$, $\lambda_{46}= -5.2784$, $\lambda_{47}= -2.2078$, $r= 0.88018$).
The stability properties are the same as in the cubic case: three negative eigenvalues ($-1.8159$, $-1.6751$ and $-0.20957$).
The Pomeron and Odderon sectors are decoupled, since the three exceptional couplings $\lambda_3, \lambda_{44}, \lambda_{45}$ vanish.
The Pomeron parameters are the same as in the pure Pomeron case at the corresponding order.
There exist three eigenvectors which span the subspace of the three  'exceptional ' couplings $\lambda_3, \lambda_{44},\lambda_{45}$. 
They have positive eigenvalues, i.e. this subspace is part of the 10-dimensional critical subspace. Inside this subspace they are orthogonal to all other 7 eigenvectors with positive eigenvalues.
Concerning the three eigenvectors with negative eigenvalues (which define the relevant directions), they are also orthogonal to three eigenvectors in the exceptional'  directions.

All this leads to the conclusion that near this fixed point the 'exceptional'  couplings define a subspace inside the critical subspace
which is orthogonal both to the remaining part of the critical subspace and to the three relevant directions. This subspace decouples from the other part. 

We observe that this special fixed point solution is associated to a critical theory conserving the Odderon number.
We do not find any other physical critical solution with all couplings being nonzero.

\begin{figure}
\begin{center}
\includegraphics[width=7cm]{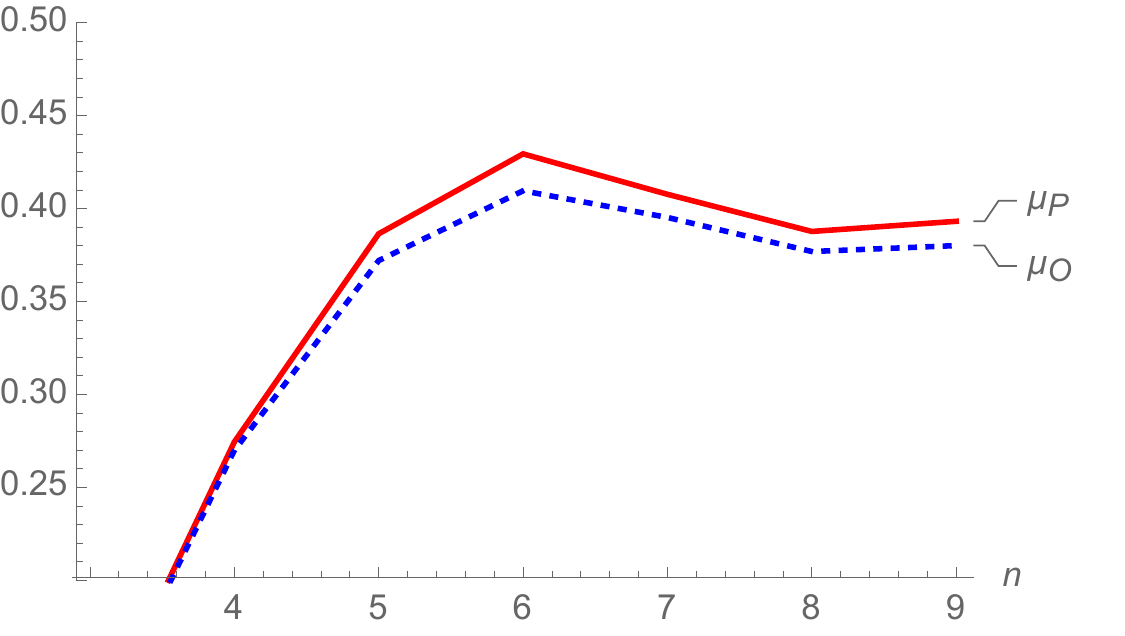}
\includegraphics[width=7cm]{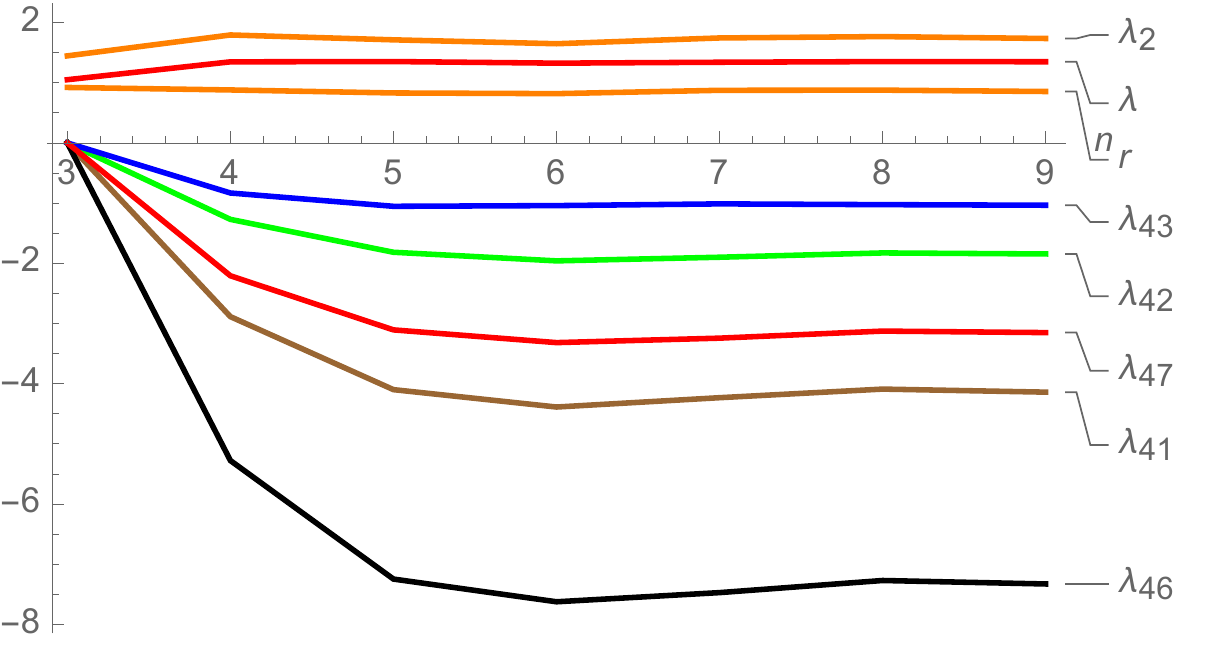}
\caption{Values of the parameters of the fixed point solution of the LPA truncations for different orders $n$ of the polynomial ($3\le n \le 9$). 
The masses (which equal intercept minus one)  $\mu_P$ (red curve) and $\mu_O$ (blue dotted curve) for the Pomeron and Odderon fields are in the left panel.
The first non zero couplings $\lambda, \lambda_2,  \lambda_{41}, \lambda_{42}, \lambda_{43}, \lambda_{46}, \lambda_{47}, r$ are reported on the right panel.}
\label{fig_trunc}
\end{center}  
\end{figure}

We then push the analysis for this special fixed point solution up to order 9 in the polynomial expansion.
We collect the results found in two figures in order to show the convergence with respect to the order of the truncation.
In Fig.~\ref{fig_trunc} we show on the left side the values for $\mu_P$ and $\mu_O$ while on the right side we give the values of the non zero couplings which characterize the truncation up to order fourth,
for all the orders $n$ between $3$ and $9$. We note that $\mu_P>\mu_O$ in all truncations. 
We see how at order $9$ a good stability is reached. We stress that all the quantities reported in this figure are not universal.

\begin{figure}
\begin{center}
\includegraphics[width=7cm]{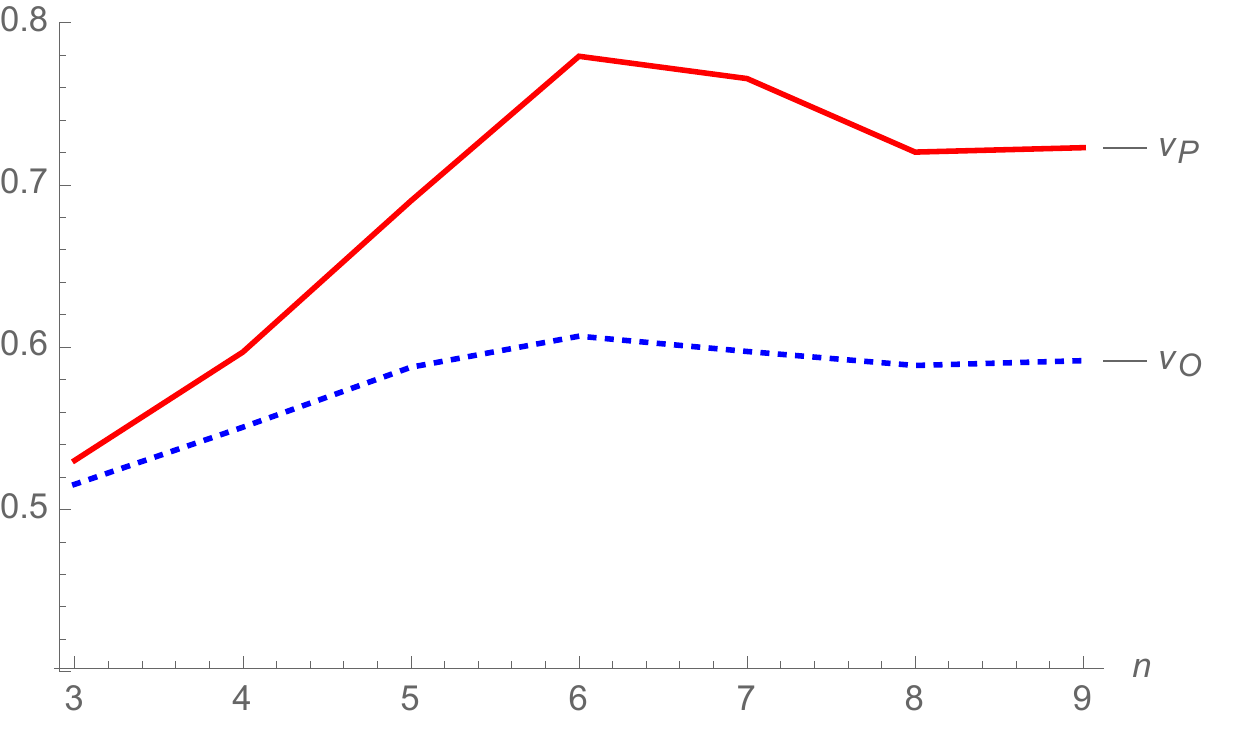}
\includegraphics[width=7cm]{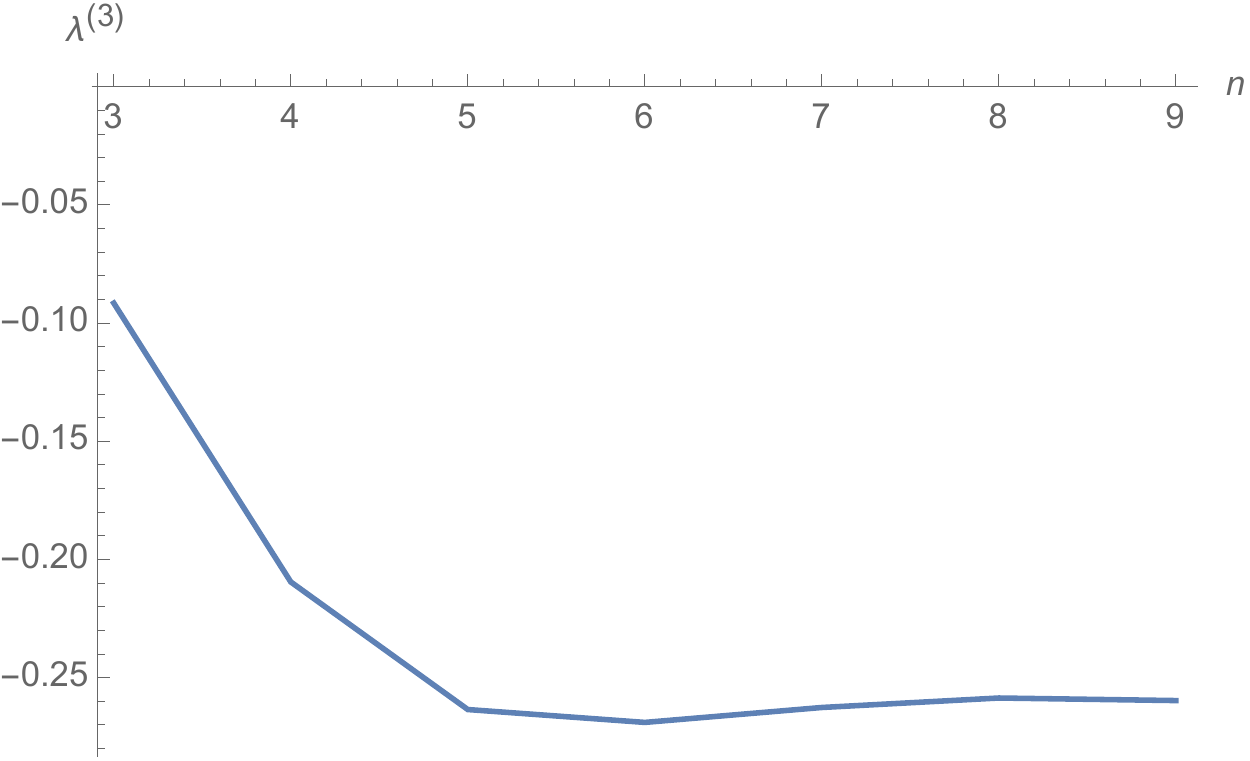}
\caption{Values of the critical exponents of the fixed point solution of the LPA truncations for different orders $n$ of the polynomial ($3\le n \le 9$). The two negative leading eigenvalues define the two critical exponents $\nu_P$ (red curve) and $\nu_O$ (blue dotted curve) for the Pomeron and Odderon fields (left panel). We report also the value of a third negative eigenvalue found in our approximation (right panel).}
\label{fig_crit}
\end{center}  
\end{figure}

In the subsequent Fig.~\ref{fig_crit} we show the critical exponents $\nu_P$ and $\nu_O$ (left plot) and the third negative eigenvalue (right plot) found at different orders of the polynomial expansion.
Also here we see that at order $9$ also the critical exponents have reached values which are almost independent of the order of the polynomial.

\subsection{The fixed point solution in continuous dimensions.}

In this last part we restrict ourselves to the lowest cubic truncation, use the expansion around the origin, include the anomalous dimension and vary the transverse dimension $D$ continuously between 0 and 4. This will provide some hints of the quality of our approximations. 
We already have the experience for the pure Pomeron sector that 
the cubic expansion is less reliable in estimating the critical exponent $\nu_P$ than an expansion around a non trivial configuration field configuration (in \cite{Bartels:2015gou} we used an expansion around the stationary point on the axes of the $(\psi,\psi^\dagger)$ plane). Since the fixed point of the interacting Pomeron-Odderon system found above leaves 
the Pomeron sector unchanged, we expect a similar situation in the present case.      
But even if we cannot expect the critical exponents $\nu_P$ and $\nu_O$ (see Fig.~\ref{fig_crit}) to be well described, it interesting to see how they connect with the result of the $\epsilon$-expansion analysis near $D=4$.

\begin{figure}
\begin{center}
\includegraphics[width=7cm]{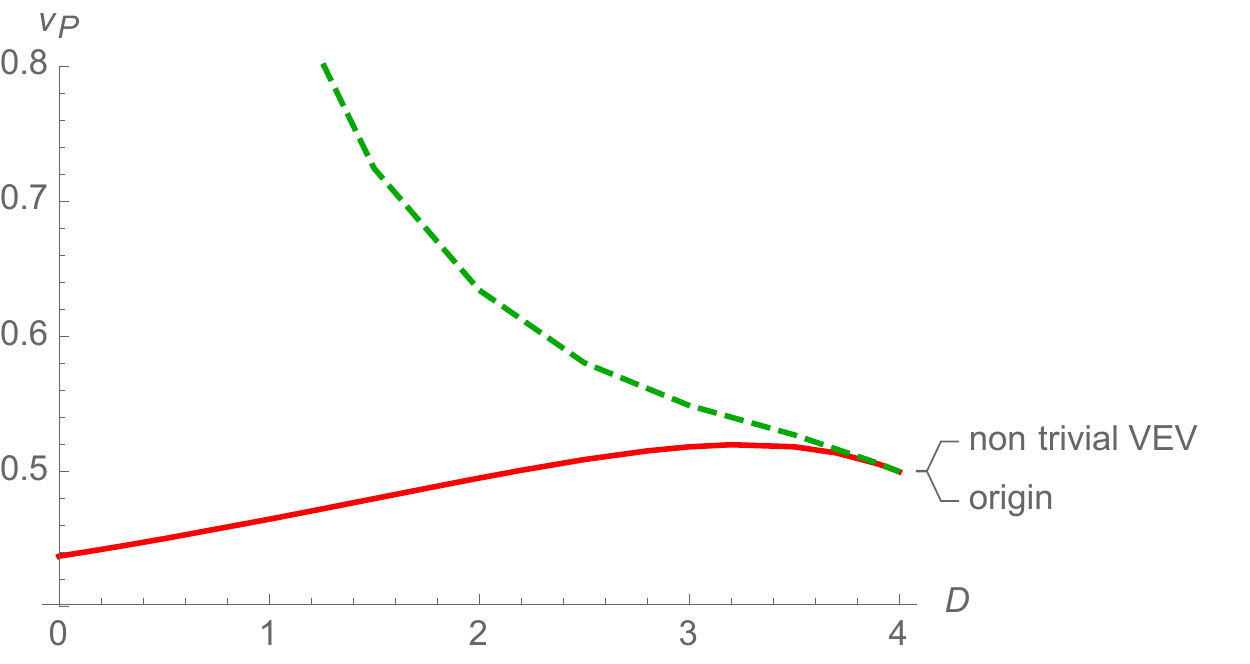}
\includegraphics[width=7cm]{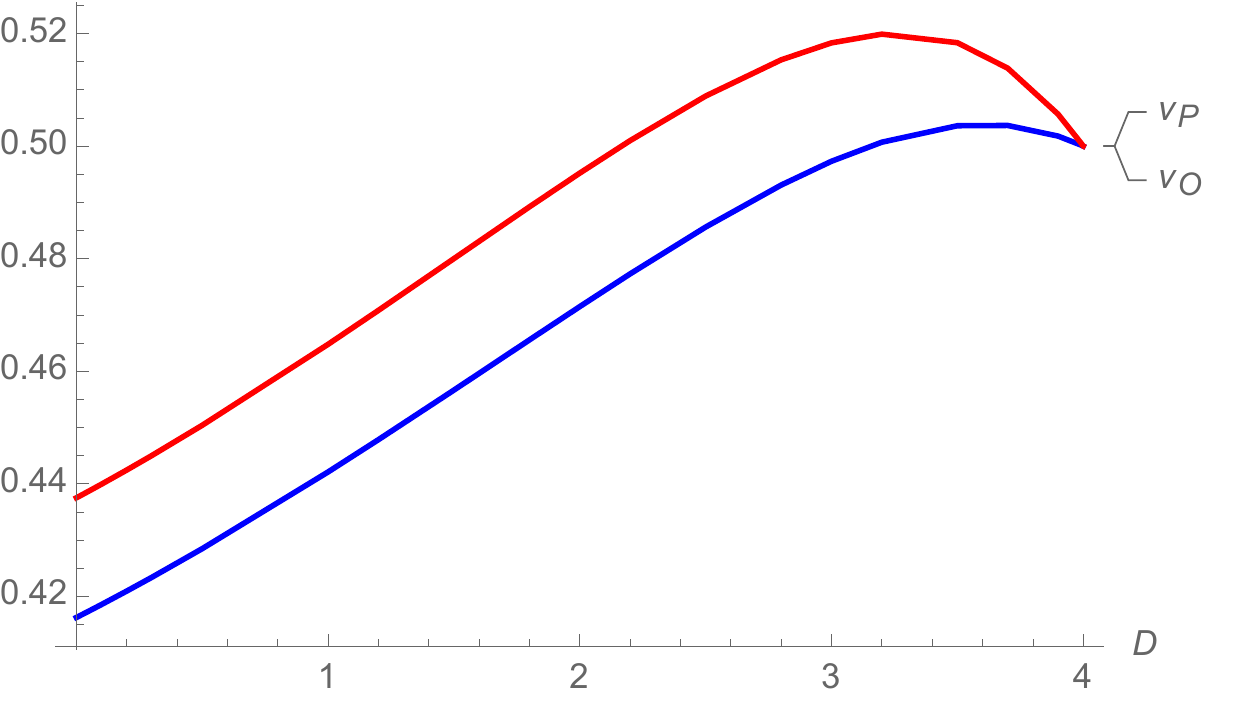}
\caption{Values of the critical exponent $\nu_P(D)$ of the pure Pomeron critical theory obtained from two different polynomial expansions (left panel). 
The values of $\nu_P$ and $\nu_O$ at the fixed point obtained from the cubic truncation around the origin for $0<D<4$.(right panel). 
In the neighborhood  of $D=4$ these curves are tangent to those associated to the $\epsilon$-expansion, according to Eq.~\eqref{nuepsilon}.}
\label{fig_DnuPO}
\end{center}  
\end{figure}

We collect some results in Fig.~\ref{fig_DnuPO} where, on the left panel we show the results of a numerical analysis for $\nu_P(D)$
for the Pomeron sector only in the two different expansions around the origin (continuous red line)
and around a non trivial stationary point on the axes (dashed green line),
while on the right panel we compare the results of the expansion around the origin $\nu_P$ and $\nu_O$.

\begin{figure}
\begin{center}
\includegraphics[width=6cm]{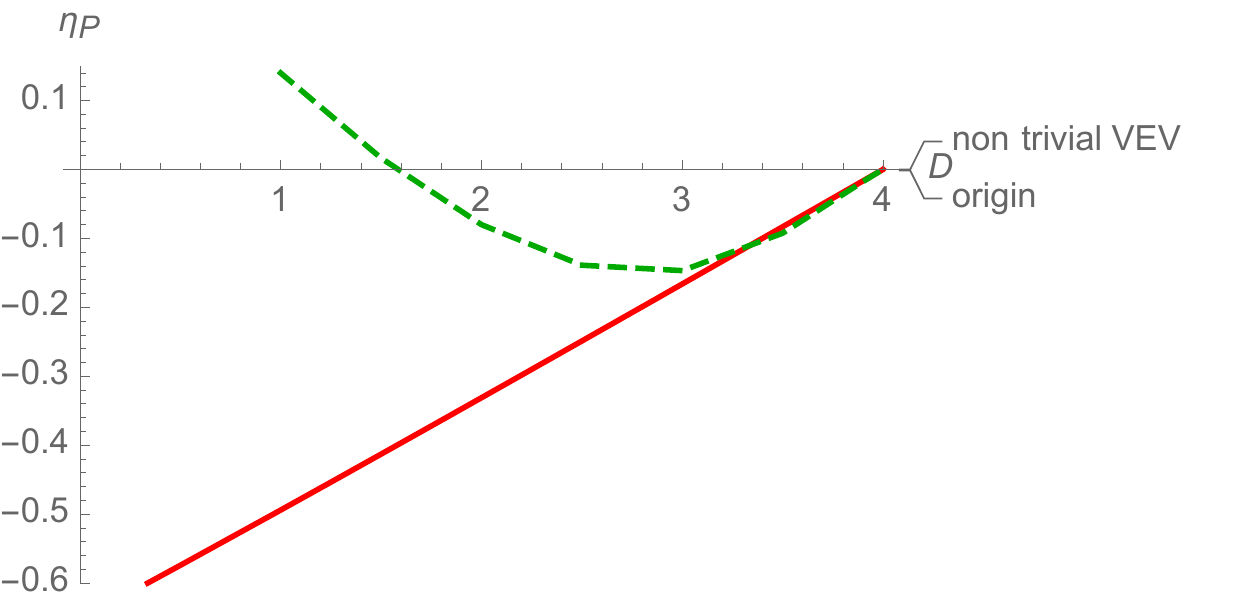}
\includegraphics[width=4cm]{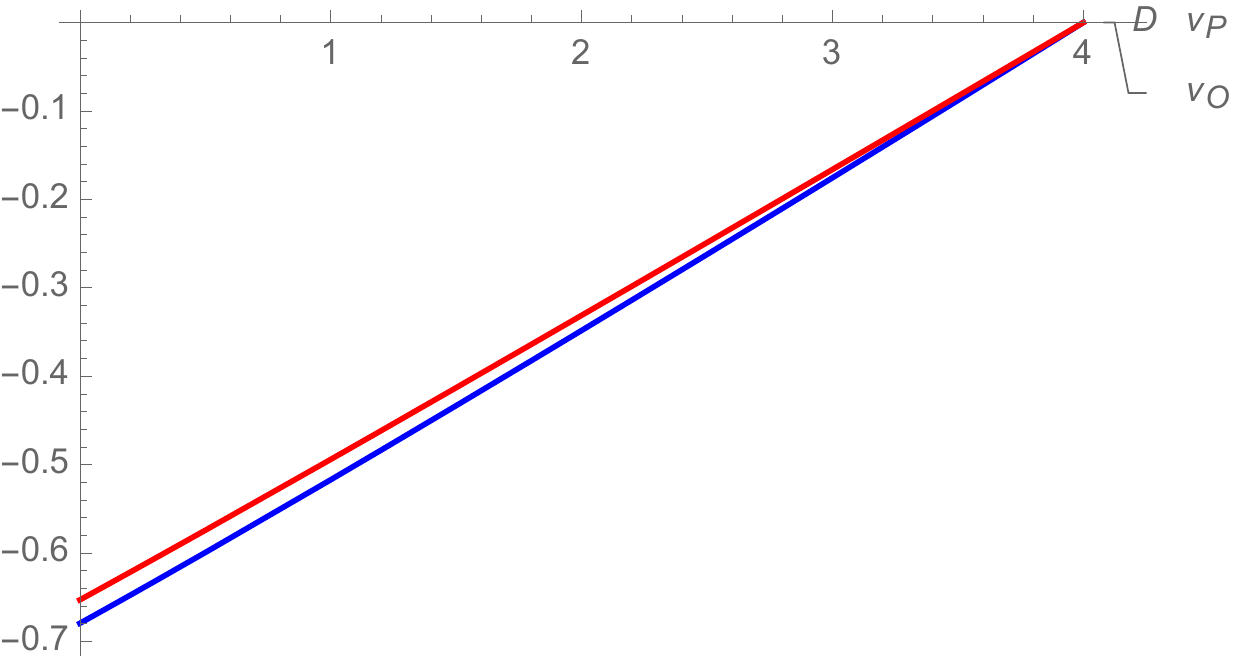}
\includegraphics[width=4cm]{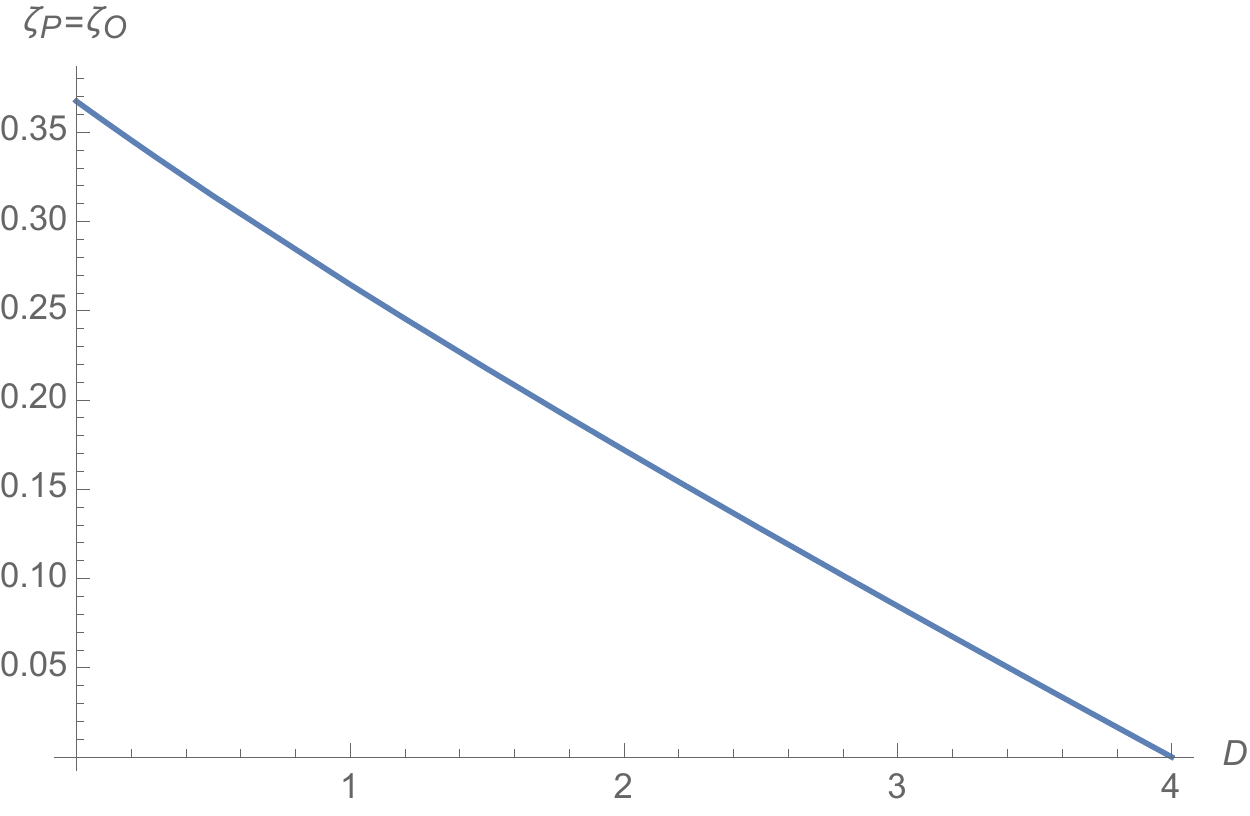}
\caption{Values of the anomalous dimension $\eta_P(D)$ of the pure Pomeron critical theory obtained from two different polynomial expansions (left panel).
Values of $\eta_P$ (red curve) and $\eta_O$ (blue dotted curve) (center panel) and of $\zeta_P=\zeta_O$ (right panel) at the fixed point obtained with a cubic truncation for $0<D<4$.
In the neighborhood of $D=4$ these curves are tangent to those associated to the $\epsilon$-expansion, according to Eq.~\eqref{specialFPepsilon}.}
\label{fig_DanomPO}
\end{center}  
\end{figure}

From our previous analysis of the pure Pomeron sector we could observe that, contrary to the critical exponent $\nu$, an expansion around the origin within the cubic truncation was able to give not too bad numerical predictions for the anomalous dimensions at $D=2$. This was not true anymore for higher order truncations. The expansion around a non trivial configuration on the axes was behaving much better at a generic order of the polynomial, even if the simple cubic truncation around the origin was giving better values. This is shown on the left panel of Fig.~\eqref{fig_DanomPO}, noting that the Monte Carlo results for a Directed Percolation model in $D=2$ which lies in the same universality class of the Pomeron RFT are pointing to a value for the anomalous dimension $\eta_P\simeq -0.4$. In the center and right plots of Fig.~\eqref{fig_DanomPO} we show $\eta_{P,O}(D)$ and $\zeta_{P,O}(D)$ respectively. They are in agreement with the behavior close to $D=4$ obtained from the $\epsilon$-expansion analysis in Eq.~\eqref{specialFPepsilon}. Finally we find that
$r(D)\simeq 0.9$  in the whole range of dimensions.

\subsection{Summary of our numerical results}
   
As the main result,  we have found a fixed point with two (or three) relevant directions: these directions are UV stable (i.e. IR unstable), whereas all other eigenvalues belong to infrared stable directions. In particular, we have the following estimates
for the critical exponents: $\nu_P\simeq 0.73$, $\nu_O\simeq 0.6$. In our approximation we find also a third negative eigenvalue $\lambda^{(3)}\simeq -0.26$ (relevant direction); 
for the anomalous dimensions we find $\eta_P\simeq -0.33$,  $\eta_O\simeq -0.35$ and $\zeta_P = \zeta_O\simeq+0.17$. 
More corrected values for the anomalous dimensions can be $20\%$ larger in magnitude according to what we observe from Monte Carlo analysis in the pure Pomeron sector.
This generalizes the previous results obtained for the pure Pomeron case, where we have found a fixed point with one relevant direction.

For such a fixed point, at first sight, the situation looks as follows. In the parameter space of the effective potential, the relevant directions define an orthogonal subspace which we name 'critical subspace'. If one starts, at $k\neq 0 $, within this subspace one ends up, for $k \to 0$, at the infrared stable fixed point. On the other hand, if one starts at a generic value outside the critical subspace (not too far away from the fixed point) the flow will eventually be attracted by the relevant direction and move away from the fixed point. 

 A closer look,  however, leads to a somewhat different picture. 
 Our fixed point analysis was done in the space of dimensionless parameters (cf. section 2), and the flow of the physical (i.e. dimensionful) parameters can be quite different.
In particular, when $k \to 0$, the nonvanishing fixed point values values of the (dimensionless) Pomeron and Odderon masses lead to vanishing physical masses, quite in the same way as 
in the pure Pomeron case discussed in \cite{Bartels:2015gou}.
A flow starting outside the critical subspace may also lead to finite values of $\mu_O$ or $\mu_P$ which can be positive or negative. Detailed features of such flows require further 
studies. Whereas for the pure Pomeron case we have performed numerical studies of the flow of the dimensionless and dimensionful parameters,
for the Pomeron-Odderon system such a study remains a task for future analysis.

There is another interesting feature of the fixed point which we have found. 
Namely, a particular set of Pomeron-Odderon couplings, although allowed by signature conservation,  vanishes at the fixed point. We interpret this result as a particular conservation law which is valid at the fixed point: $t$-channel states formed by Pomeron and Odderons conserve the number of Odderon pairs. In particular, in the critical regime there are no transitions from pure  Pomeron states to states containing Odderons.

\section{Discussion and outlook} 

In this paper we have extended our previous fixed point analysis of Pomeron reggeon field theory to a system of interacting Pomeron and Odderon fields in the infrared limit. 
Let us briefly discuss the potential implications of our results for physical processes.
To relate our flow analysis to the high energy behavior of physical scattering amplitude we will
assume that the infrared momentum cutoff, $k$, is related to the transverse extension of the scattering system, the radius $R(s)$. Since $R$ grows with energy, we expect $k^2 \sim 1/R^2$ to go to zero, for example proportional to $1/\ln s$ (neglecting anomalous dimension).        

As discussed in the introduction,  the main motivation for including the Odderon comes from the observation that, in the UV region where perturbative QCD applies, there exist two fundamental composite states of reggeized gluons, the BFKL Pomeron with intercept well above one and a very small slope, and the Odderon with intercept at (or very close to) one and a small slope. 
This raises the question, when moving towards the nonperturbative IR region, to what extent the interactions between these fundamental fields lead to serious modifcations, e.g. a suppression of the Odderon exchange at high energies.

In our fixed point analysis we have found, very similar to the pure Pomeron case, a fixed point which is infrared stable in all but two (possibly three) directions. These 'relevant' directions define an orthogonal 'critical subspace'. This fixed point structure allows for several asymptotic solutions.
 
If we start, in the UV region, inside the critical subspace, we end up, in the IR limit, at the fixed point. At this fixed point, both the Pomeron and the Odderon have intercept one.
From (\ref{dimensional}) we see that near the fixed point both intercepts, $\alpha_P(0)-1 =\mu_P/Z_P \sim k^{2-\zeta}  \tilde{\mu}_P$ and
$\alpha_O(0)-1 =\mu_O/Z_O\sim k^{2-\zeta} \tilde{\mu}_O$ go to zero as $k$ becomes small. Since the fixed point value of $ \tilde{\mu}_P$ is slightly larger than $\tilde{\mu}_O$ we conclude that, for small but nonzero values of $k$, the Pomeron intercept is larger than the Odderon intercept. This is consistent with the Pomeranchuk theorem. Moreover, a first study of the flow equations (in the cubic truncation)  further away from the fixed point shows that most trajectories belong to $\mu_P$ above and $\mu_O$ below its 
fixed point values. This is consistent with our expectations for the starting points in the UV region; the Pomeron 
value $\mu_P$ should be positive, whereas the Odderon mass $\mu_O$ shoul be at (or close to)
zero.  
However, the most important conclusion to be drawn from this  fixed point analysis is that the Odderon exists in the IR limit and does not die out with energy.  
It should be clear that our study does not include the couplings of the Pomeron and Odderon to 
external particles; phenomenological studies indicate the Odderon couplings are smaller then
those of the Pomeron. This may explain the smallness of the Odderon  contribution.    
For the ratio of the Odderon and Pomeron slopes we find a fixed point value slightly below one.    
Phenomenologically, not much is known about the Odderon slope \cite{Ewerz:2003xi,Dosch:2002ai}, and our result might be seen as an asymptotic  prediction.         

At this stage, however, we also have to allow for another possibility. If the UV starting point lies 
outside the critical subspace, the flow may lead to another situation where the intercepts may still be finite, but can be above one (supercritical), at one (critical) or below one (subcritical).
These solutions have to be investigated seperately.  

As mentioned before, at the fixed point those Pomeron-Odderon couplings which change the number of Odderon pairs 
are zero. In particular, the Pomeron $\to$ two Odderon coupling $POO$ (which was found to be nonzero in pQCD) vanishes. Thus the Pomeron is not affected by the presence of the Odderon. The Odderon, on the other hand, undergoes nonzero absorption by the Pomeron.        
 
The possibility that in the deep IR region the $POO$ vertex is suppressed may also have phenomenological consequences.
Processes involving a simple Odderon exchange, like hadron scattering $pp$ - $p\bar{p}$ 
or meson photo-production~\cite{Bartels:2001hw} would be allowed in asymptotic IR, 
while high mass diffractive processes with a $POO$ vertex would be possibly allowed in an intermediate (more perturbative) regime ~\cite{Bartels:2003zu},
but suppressed in the deep IR region.

Like the Pomeron RFT, the extended RFT model studied in this paper may be related to a generalized multicomponent directed percolating system, characterized by some special symmetries. For the latter we have found slightly different critical exponents which suggest the existence of a new universality class.  This is certainly true in the vicinity of $D=4$, from the $\epsilon$-expansion analysis. Nevertheless more refined analysis employing larger truncations should be done for the case of two transverse dimensions.

There are several questions to be addressed by future studies. First, we have to search for possible alternative fixed points. For this goal our fixed point analysis
has to be improved by considering polynomial expansions around stationary points away from the origin. 
Experience from the pure Pomeron case has shown that such expansions seem to have better convergence properties when increasing the order of truncation.  

Most important, however, is the next - essential - step in our program, the study of the 
flow equations formulated in the  region of perturbative QCD. 
In this region, one of the crucial features to be addressed within the framework of the flow equations is the fact that the BFKL Pomeron and also the Odderon are a composite states of reggeized gluons, i.e. we have to use a formulation which includes both the reggeized gluon as the fundamental field and the (nonlocal) color singlet composite fields. 
Work along these lines is in progress.

\section*{Acknowledgements}
Part of this work has been done while one of us (J.B.) has been visiting the Departamento de Fisica, Universidad Tecnica Santa Maria, Valparaiso, Chile, and the hospitality is gratefully acknowledged. This research was supported  by the Fondecyt (Chile) grants 1140842,  MEC - 80140070 and by DGIP/USM.

\appendix
\section{Stationary points}

Let us search for stationary points of the potential.
Non trivial stationary points (saddles or extrema) may indeed provide a better field configuration around which one can perform a polynomial expansion of the potential, with respect to the speed of convergence properties, as was observed in the pure Pomeron RFT analysis.
We first will restrict ourselves to the lowest  truncation. 

The first derivatives are:
\ba
\label{V-psidag}
\frac{\partial V}{\partial \psidag} &=& \psi \left( -\mu_P  +i\lambda (2 \psidag +\psi)\right)  + \left( i\lambda_2\chidag  + \lambda_3 \chi\right) \chi \\
\label{V-psi}
\frac{\partial V}{\partial \psi} &=& \psidag \left  (-\mu_P  +i\lambda (2 \psi +\psidag)\right) +\chidag \left( i\lambda_3\chi + \lambda_3 \chidag\right)\\ 
\label{V-chidag}
\frac{\partial V}{\partial \chidag} &=& \chi\left( -\mu_O   +i\lambda_2 (\psidag +\psi) \right) +2 \lambda_3\chidag \psi  \\
\label{V-chi}
\frac{\partial V}{\partial \chi} &=&\chidag \left(  -\mu_O   +i\lambda_2  ( \psi +\psidag) \right) +2 \lambda_3\psidag\chi.
\ea
 The last two equations are linear in $\chi$ and $\chidag$:
either we have 
\be
(\chi,\chidag)=(0,0)
\ee
or the determinant vanishes:
\be
\label{det}
\left(\mu_O - i\lambda_2(\psi+\psidag) \right)^2 - 4 \lambda_3^2 \psi \psidag =0.
\ee  
In first case we are back to the pure Pomeron case with the four stationary points
\be 
(\psi,\psidag)= (0,0),   (\psi,\psidag)=(\frac{\mu_P}{i\lambda},0), 
(\psi,\psidag)=(0,\frac{\mu_P}{i\lambda}), (\psi,\psidag)=(\frac{\mu_P}{3 i\lambda},\frac{\mu_P}{3 i\lambda}).
\ee
In the second case we have several possibilities. First we set   
\be
\psi=0.
\ee
Then, from the condition (\ref{det}) we derive 
\be
\psidag = \frac{\mu_O}{i\lambda_2}
\ee
and from (\ref{V-chi}) and  (\ref{V-psi})
\be
\chi=0,\,\, {\chidag}^2= \frac{\mu_0}{i \lambda_2\lambda_3} \left( \mu_P -  \mu_O 
\frac{\lambda}{\lambda_2}\right).
\ee
There exists also the solution
\be
\psi=\psidag=\phi.
\ee
Eq.(\ref{det}) leads to the two solutions  
\be
\phi_{\pm} = \frac{\mu_0}{2(i\lambda_2 \pm \lambda_3)}
\ee
and (\ref{V-chidag}) (or (\ref{V-chi}))
imply either  
\be 
\chi = -\chidag
\ee
with
\be
\chi^2= \phi_+\frac{\mu_P -3i\lambda \phi_+}{\lambda_2-i\lambda_3}
\ee
or 
\be 
\chi = \chidag
\ee
with
\be
\chi^2= \phi_-\frac{\mu_P -3i\lambda \phi_-}{\lambda_2-i\lambda_3}.
\ee

There are also stationary curves (one parameter family of stationary points) of the potential, which we do not report here.


\end{document}